\documentclass[10pt]{article}
\usepackage{graphicx} 
\usepackage{amsmath} 
\usepackage{amsfonts}
\usepackage{amssymb}
\usepackage{subfigure}
\usepackage{multirow}

\title{ On the Paramagnetic Impurity Concentration of Silicate Glasses  
from Low-Temperature Physics } 
\author{ Silvia Bonfanti$^{a,b}$ and Giancarlo Jug$^{a,c}$\cite{J} \\
$^a$Dipartimento di Scienza ed Alta Tecnologia and To.Sca.Lab \\
Universit\`a dell'Insubria, Via Valleggio 11, 22100 Como, Italy \\
$^b$Laboratoire Charles Coulomb, Universit\'e de Montpellier \\
Place Eug\`ene Bataillon, F-34095 Montpellier Cedex 5 - France \\
$^c$INFN -- Sezione di Pavia, Italy and IPCF -- Sezione di Roma, Italy} 

\date{\today} 

\begin{document}

\maketitle

\begin{abstract}
The concentration of paramagnetic trace impurities in glasses can be determined 
via precise SQUID measurements of the sample's magnetization in a magnetic
field. However the existence of quasi-ordered structural inhomogeneities in the 
disordered solid causes correlated tunneling currents that can contribute to 
the magnetization, surprisingly, also at the higher temperatures. We show that 
taking into account such tunneling systems gives rise to a good agreement 
between the concentrations extracted from SQUID magnetization and those 
extracted from low-temperature heat capacity measurements. Without suitable 
inclusion of such magnetization contribution from the tunneling currents we 
find that the concentration of paramagnetic impurities gets considerably 
over-estimated. This analysis represents a further positive test for the 
structural inhomogeneity theory of the magnetic effects in the cold glasses. 
 \end{abstract}


\section{Introduction}
Multi-component silicate glasses are technologically important insulating 
structural materials, usually containing trace paramagnetic impurities 
(typically Fe$^{2+}$ and Fe$^{3+}$ from the fabrication process). Such dilute 
impurities give rise to nearly-ideal Langevin paramagnetism that can be 
exploited in low-temperature thermometry~\cite{Her2000}. A by-product of the 
magnetization measurements (usually through SQUID magnetometry) is the 
determination of the impurity concentration (in the atomic ppm region). For 
example: commercial borosilicate glasses BK7 (optical glass) and Schott's 
Duran (laboratory glass) have (reportedly~\cite{Her2000,Sie2001,Lud2003}) 
about 6 ppm and 120 ppm (or 180 ppm~\cite{Her2000}) of diluted iron 
impurities, respectively, according to SQUID-magnetometry.

Knowledge of such paramagnetic impurity concentration is important also where 
the fundamental physics of disordered solids is concerned. At low temperatures 
(below 1 K, normally) the physics of a disordered solid is known to be 
dominated by low-energy excitations that go under the name of tunneling systems 
(TSs)~\cite{Esq1998}. These TSs are dynamical defects that give rise to 
quasi-universal physical properties that can also be exploited in 
low-temperature thermometry~\cite{Sch1994}. A celebrated example is the excess 
TS contribution to the real part $\epsilon'$ of the dielectric constant at low 
frequency, which depends logarithmically on the temperature 
$T$~\cite{Phi1987,Car1994,Jug2013}. After 40 years of research, however, the 
precise microscopic nature of the TSs is still a mystery. More recently, this 
dielectric constant $\epsilon'$ has been discovered to be sensitive to weak 
magnetic fields ($B\sim$~10$^2$ to 10$^3$ G)~\cite{Woh2001} for some 
multi-component silicate glasses and possible explanations for this unexpected 
magnetic effect (observed also in other physical 
properties~\cite{Sie2001,Lud2003}) that have been proposed involve nuclear 
quadrupole moments~\cite{Wue2002}, structural inhomogeneities~\cite{Jug2004} 
and also paramagnetic impurities~\cite{Bor2007}. 
Therefore, a precise determination of the impurity concentration is essential 
also in order to decide among different explanations for the magnetic effects. 

Research on the ultimate nature of the TSs is, moreover, receiving renewed 
interest in view of the fact that the TSs have been recognized to be the 
cause of decoherence in Josephson-junction based quantum computing devices 
(the tunneling JJ barrier being typically amorphous)~\cite{Sim2004}. Moreover, 
studies of aging~\cite{Ami2012} in glasses (hard and polymeric)~\cite{Osh2003} 
at very low temperatures depend on a more complete description of the physics 
of the TSs. The issue of the origin of the magnetic effects therefore helps 
in improving knowledge about the nature of the TSs so as to 
minimize~\cite{Pai2010,Liu2014} (or exploit~\cite{Zag2006}) their decoherence
(or coherence) effect in SCJJ qubits. The study of the structure of real glasses 
at low temperatures provides, in turn, alternative information on the mechanism 
for the glass transition (normally investigated from the liquid 
state~\cite{Don2001,Ber2011}).  

In this paper we investigate mainly the issue of the determination via SQUID 
magnetization of the concentration $n_J$ of paramagnetic 
impurities~\cite{Her2000}, also in view of the fact that the theoretical 
analysis~\cite{Jug2004} of the magnetic effect~\cite{Sie2001,Ste1976} in the 
heat capacity $C_p$ of some multi-silicate glasses produced values of $n_J$ 
systematically much lower than those quoted in the 
literature~\cite{Her2000,Sie2001,Lud2003,Woh2001} (and obtained from SQUID 
measurements). We briefly review the analysis of some $C_p(T,B)$ data in the 
range 0.6 to 1.3 K~\cite{Sie2001}, then apply our model to the calculation of 
the TS contribution to the magnetization $M$ and analyze the available 
magnetization $M(T,B)$ data in the range 4 to 300 K
~\cite{Her2000,Sie2001,Lud2003} with our formula added to Langevin's 
contribution from the paramagnetic impurities. We find that the concentrations 
$n_J$ of such impurities extracted from both types of measurements will get
to agree with each other only when the TS contributions (from our model, e.g.) 
to both $C_p(T,B)$ and $M(T,B)$ get to be added to Langevin's known 
expressions for the paramagnetic impurities' contributions. We apply our 
analysis to available data for the borosilicate Duran glass and for the 
multi-silicate glass of composition Al$_2$O$_3$-BaO-SiO$_2$ (in short AlBaSiO, 
or BAS, reported concentration $\bar{n}_{Fe}\simeq$~100 ppm
~\cite{Sie2001,Lud2003,Woh2001}) since these glasses have shown the most 
remarkable magnetic effects~\cite{Lud2003}. The concentrations $n_J$ that 
we find for the Fe impurities are, however, about 60 to 80\% lower than those 
quoted in the literature for these glasses, so the correct description of the 
magnetic-sensitive TSs becomes important also for the applications of 
low-temperature physics.  
        
This paper is organized as follows: in Section 2 we present the foundations
of the extended tunneling model that we use, then in Section 3 we apply the
model to the explanation of the magnetic effect in the $C_p(T,B)$ data. In
Section 4 we derive the contribution from the TSs in our model to the 
magnetization $M(T,B)$ and analyze the available data with our formula. The 
values of $n_J$ extracted from both quantities $C_p$ and $M$ are compared in 
Section 5, which also contains our conclusions. In the Appendix we present
some preliminary results for the analysis of some measurements in BK7 glass.
\section{The Extended Tunneling Model}
The modern justification for the tunneling model is based on the belief that
glasses at sufficiently low temperatures are characterized by a potential 
energy landscape (PEL), that can be investigated only by means of classical 
computer simulations of atomic configurations. Out of the many local minima 
of the PEL some local potentials, normally thought to be double-welled (DWP), 
give rise to TSs, normally replaced by two-level systems (2LSs) that at low 
temperatures are characterized by the tunneling Hamiltonian~\cite{Phi1987}:
\begin{equation}
{\cal H}_{2LS}=-\frac{1}{2}\left( \begin{array}{cc}
\Delta & \Delta_0 \cr
\Delta_0 & -\Delta \end{array} \right).
\label{2lstunneling}
\end{equation}
Here the parameters $\Delta$ (the energy asymmetry) and $\Delta_0$ (twice the 
tunneling parameter) are typically characterized by a probability 
distribution that views $\Delta$ and $\ln(\Delta_0)$ (the latter linked to 
the DWP energy barrier) broadly (in fact uniformly) distributed throughout
the disordered solid~\cite{Phi1987}:
\begin{equation}
{\cal P}_{2LS}(\Delta,\Delta_0)=\frac{\bar{P}}{\Delta_0}
\label{2lsdistribution}
\end{equation}
where some cutoffs are introduced when needed and where $\bar{P}$ is a 
material-dependent parameter, like the cutoffs. In this justification of the 
TSs clearly the tunneling ``particle'' cannot be typically a real atom/ion of 
the glass, but rather an effective, fictitious particle: jumps between 
contiguous low-lying minima of the PEL correspond to the rearrangements of 
several atoms/ions, if not large parts of the entire atomic configuration. 

Much progress has been done in understanding the physics of glasses at low 
temperatures with the help of the above simple model (the standard tunneling 
model, STM)~\cite{Esq1998}, which has however important limitations. The 
recently discovered magnetic effects in non-magnetic 
glasses~\cite{Sie2001,Lud2003,Woh2001,Wue2002}, for example, 
cannot be explained without a suitable extension of the STM. Our own extension 
brings three realistic considerations in the modeling of the structure of 
real glasses. The first is that glasses can be no longer considered fully 
homogeneously disordered solids at the intermediate atomic scales, for there 
is mounting experimental evidence 
that the structure is spatially inhomogeneous with ``better ordered'' regions 
being inter-dispersed in an otherwise featureless homogeneously disordered 
matrix. One way to look at these regions of enhanced atomic ordering is that 
they are the thermal-history continuation to temperatures below the glass 
transition temperature $T_g$~\cite{Vol2005} of the slower-particles regions 
present, within the sea of faster-particle regions, in the 
dynamical-heterogeneity picture~\cite{Bir2011} of the supercooled liquid 
phase, between $T_g$ and the melting temperature $T_m>T_g$. 
These slower-particle regions are in fact also better ordered, as expected.
We name these regions in the glassy phase {\it regions of enhanced regularity} 
(RERs), but some other names have been proposed in the literature: 
{\it cybotactic groupings} from a critical analysis of the X-ray and neutron 
scattering data in amorphous solids~\cite{Wri2014}, {\it para-crystals} from 
combined electron-diffraction and fluctuation electron microscopy of {\it a}-Si 
films~\cite{Tre2012}. Similar conclusions about partial devitrification have 
been reported for the metallic glass Zr$_{50}$Cu$_{45}$Al$_5$ using combined 
Monte Carlo simulation and fluctuation electron microscopy~\cite{Hwa2012}. 
Besides, the amorphous solids of general composition 
(MgO)$_x$(Al$_2$O$_3$)$_y$(SiO$_2$)$_{1-x-y}$ are termed {\it ceramic glasses} 
and are even known to contain embedded micro-crystals~\cite{Bac1999}. 
The evidence from X-ray analysis for crystalline-like ordering in quenched 
network glasses is in fact most compelling in the case of multi-component 
materials~\cite{Wri2014}. For these glasses the distribution 
(\ref{2lsdistribution}) of the 2LS parameters should be partially abandoned 
in favour of a different distribution, for a subsystem of TSs nesting within 
the RERs, having a form favouring the near-symmetry of the local wells of 
the DWP, e.g.:
\begin{equation}
{\cal P}_{2LS}^{\ast}(\Delta^{\ast},\Delta_0^{\ast})=
\frac{P^{\ast}}{\Delta_0^{\ast}\Delta^{\ast}}
\label{nd2lsdistribution}
\end{equation}
where the parameters $\Delta^{\ast}, \Delta_0^{\ast}$ and $P^{\ast}$ 
now refer to that subsystem. One should then use at the very 
least~\cite{Jug2004} a collection of 
the two types of TSs described above when dealing with multi-component glasses 
(for which the strongest magnetic effects have been observed).
\begin{figure}[!Hbtp]
  \centering
  \includegraphics[scale=0.30] {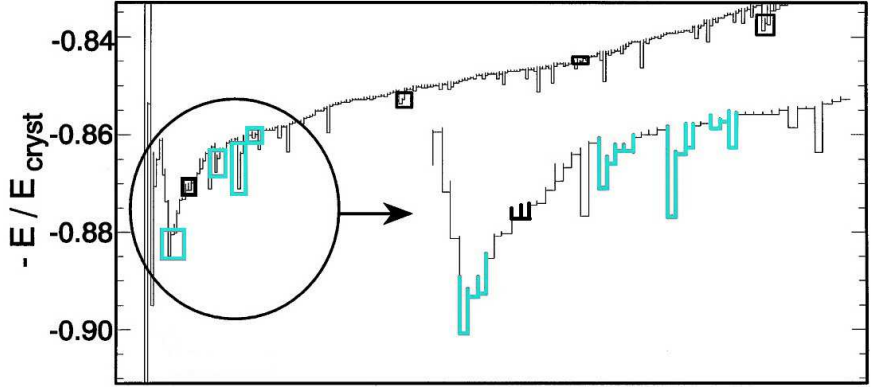}
\caption{The PEL for a BLJ mixture of 32 particles, mapped in 1D according to 
the procedure by Heuer. Highlighted in black are the 2LSs, then in blue the 
3LSs, 4LSs, ... ensuing from multi-welled local potentials. Adapted from 
\cite{Heu1997}}
  \label{landscape}
\end{figure}
The second consideration is that the computer-generated PEL typically contains 
more complicated, local multi-welled potentials (MWPs) as well as DWPs 
(indeed, a representation of the PEL solely by means of DWPs and 2LSs seems 
rather oversimplified, though of obvious practical theoretical 
advantage~\cite{Yu1988}). In Fig.~\ref{landscape} the one-dimensional map of 
a molecular dynamics-generated PEL for a system of 32 particles interacting 
through a binary mixture Lennard-Jones potential in the glassy phase has been 
reproduced~\cite{Heu1997}. There are clearly DWPs, but also MWPs as we have 
highlighted. The tunneling Hamiltonian of a MWPs is easily written 
down as~\cite{Jug2004}:
\begin{equation}
{\cal H}_{3LS}=\left( \begin{array}{ccc}
E_1 & D_0 & D_0 \cr
D_0 & E_2 & D_0 \cr
D_0 & D_0 & E_3 \end{array} \right)
\label{3lstunneling}
\end{equation}
where $E_1, E_2, E_3$ are the energy asymmetries between the wells (we have 
chosen the simplest case, having three wells) and $D_0$ is 
the most relevant tunneling amplitude (through saddles of the PEL, in fact). 
This 3LS Hamiltonian has the advantage of readily allowing for the inclusion 
of a magnetic field $B>0$, when coupling orbitally with a tunneling 
``particle'' having charge $q$ ($q$ being some multiple of the electron's 
charge $-e$)~\cite{Jug2004}: 
\begin{equation}
{\cal H}_{3LS}(B)=\left( \begin{array}{ccc}
E_1 & D_0e^{i\varphi/3} & D_0e^{-i\varphi/3} \cr
D_0e^{-i\varphi/3} & E_2 & D_0e^{i\varphi/3} \cr
D_0e^{i\varphi/3} & D_0e^{-i\varphi/3} & E_3 \end{array} \right)
\label{3lsmagtunneling}
\end{equation}
where $\varphi/3$ is the Peierls phase for the tunneling particle through
a saddle in the field, and $\varphi$ is the Aharonov-Bohm phase for a 
tunneling loop and is given by the usual formula:
\begin{equation}  
\varphi=2\pi\frac{\Phi}{\Phi_0}, \qquad \Phi_0=\frac{h}{\vert q\vert}
\label{ABphase}
\end{equation}  
$\Phi_0$ being the appropriate flux quantum ($h$ is Planck's constant) and 
$\Phi={\bf B}\cdot{\bf S}$ the magnetic flux threading the area $S$ formed 
by the tunneling paths of the particle in this simple model. The energy 
asymmetries $E_1, E_2, E_3$ typically enter through their combination 
$D\equiv\sqrt{E_1^2+E_2^2+E_3^2}$. One can easily convince oneself that if 
such a MWP is used with the standard parameter distribution,
Eq.~(\ref{2lsdistribution}) with $D, D_0$ replacing $\Delta, \Delta_0$, for
the description of the TS, one would then obtain essentially the same physics 
as for the STM 2LS-description. In other words, there is no need to complicate
the minimal 2LS-description in order to study glasses at low temperatures,
unless structural inhomogeneities of the RER-type and a magnetic field are
present. Without the RERs, hence no distribution of the type 
(\ref{nd2lsdistribution}), the interference from separate tunneling paths is
only likely to give rise to a very weak Aharonov-Bohm effect. Hence, it will be 
those TSs nesting within the RERs that will give rise to an enhanced A-B effect 
and these TSs can be minimally described -- for example -- through Hamiltonian 
(\ref{3lsmagtunneling}) and with distribution (\ref{nd2lsdistribution}) thus 
modified to favour near-degeneracy~\cite{Jug2004}:
\begin{equation}
{\cal P}_{3LS}^*(E_1,E_2,E_3;D_0)=\frac{P^{\ast}}{D_0(E_1^2+E_2^2+E_3^2)}.
\label{atsdistribution}
\end{equation}
We remark that the incipient ``crystallinity'' of the RERs calls for 
near-degeneracy in $E_1, E_2, E_3$ simultaneously and not in a single one 
of them, hence the correlated form of (\ref{atsdistribution}). Other
descriptions, with four-welled potentials or modified three-dimensional DWPs 
are possible for the TSs nested in the RERs and lead to the same physics as 
from Eqs.~(\ref{3lsmagtunneling}) and (\ref{atsdistribution}) 
above~\cite{Bon2015} (which describe what we call the anomalous tunneling 
systems, or ATSs, nesting within the RERs).

The final and most important consideration is that the TSs appear to be rather 
diluted defects in the glass (indeed their concentration is of the order of
magnitude of that for trace paramagnetic impurities, as we shall see), hence 
the tunneling ``particles'' are embedded in a medium otherwise characterized 
only by simple acoustic-phonon degrees of freedom. This embedding, however, 
means that the rest of the material takes a part in the making of the 
tunneling potential for the TS's ``particle'', which itself is not moving 
quantum-mechanically in a vacuum. Sussmann~\cite{Sus1962} 
has shown that this leads to local trapping potentials that (for the case of 
triangular and tetrahedral perfect symmetry) must be characterized by a 
degenerate ground state. This means that, as a consequence of this TS 
embedding, our minimal model (\ref{3lsmagtunneling}) must be chosen with a 
positive tunneling parameter~\cite{Jug2004}:
\begin{equation}
D_0>0
\label{degeneracy}
\end{equation}
where of course perfect degeneracy is always removed by weak disorder in the 
asymmetries. The intrinsic near-degeneracy of (\ref{atsdistribution}) implies 
that this model should be used in its $D/D_0\ll 1$ limit, which in turn reduces 
the ATSs to effective magnetic-field dependent 2LSs and greatly simplifies the 
analysis together with the limit $\varphi\to 0$ which we always take for 
relatively weak magnetic fields. Our extended tunneling model (ETM) consists 
then in a collection of independent, non-interacting 2LSs described by the STM 
and 3LSs described by Eqs.~(\ref{3lsmagtunneling}) and (\ref{atsdistribution}) 
above in the said $D/D_0\ll 1$ and $\varphi\to 0$ limits, the 3LSs nested 
within the RERs and the magnetic-field insensitive 2LSs distributed in the 
remaining homogeneously-disordered matrix.
\begin{figure}[!h]
  \centering
  \includegraphics[scale=0.60] {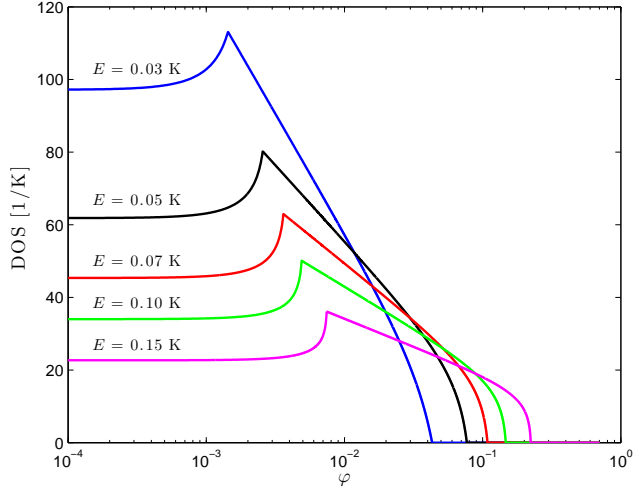}
\caption{The magnetic-sensitive part of the density of states (DOS) as a
function of the A-B phase $\varphi$ (proportional to the magnetic field $B$)
and different energies ($n_{ATS}P^{\ast}$ has been set to 1). The shape of 
this part of the DOS (coming from the MWPs with a parameter distribution 
(\ref{atsdistribution}) favouring near-degeneracy) is the ultimate source of 
all the magnetic effects. 
The cusp is an artifact of the effective 2LS approximation~\cite{Jug2004}, but
also of the existence of upper and lower bounds for $D_0$ owing to the nature
of the RER atomic structure.}
  \label{dosfig}
\end{figure}
Our ETM has been able to explain the magnetic effects in the heat 
capacity~\cite{Jug2004}, in the real~\cite{Jug2009} and 
imaginary~\cite{Jug2014} parts of the dielectric constant and in the 
polarization echo amplitude~\cite{Jug2014} measurements reported to date for 
various glasses at low temperatures, as well as the composition-dependent 
anomalies~\cite{Jug2013,Jug2010}. 
The new physics is provided by the magnetic-field dependent TS density of 
states (DOS) which acquires a term due to the near-degenerate 
MWPs~\cite{Jug2004} that gets added up to the (nearly) constant DOS from the 
STM 2LSs (having density $n_{2LS})$:
\begin{equation}
g_{tot}(E,B)=n_{2LS}\bar{P}+n_{ATS}\frac{P^{\ast}}{E}f_{ATS}(E,B)\theta(E-E_c)
\label{dos}
\end{equation}
where $n_{ATS}$ is the ATSs' concentration,
$f_{ATS}$ is a magnetic-field dependent dimensionless function, already 
described in previous papers~\cite{Jug2004}, and $E_c$ is a material and 
$B$-dependent cutoff. 
The $1/E$ dependence is a consequence of the chosen tunneling parameter 
distribution, Eq.~(\ref{atsdistribution}) (or, to the same effect, 
Eq.~(\ref{nd2lsdistribution})), and gives rise to a peak in $g_{tot}$ near 
$E_c$ that is rapidly eroded away as soon as a weak magnetic field is 
switched on. The form and evolution of the magnetic part of the DOS is shown 
in Fig.~\ref{dosfig} for some typical parameters, as a function of $B$ for 
different values of $E$. 
This behaviour of the DOS with $B$ is, essentially, the underlying mechanism 
for all of the experimentally observed magnetic field effects in the cold 
glasses within this
model: the measured physical properties are convolutions of this DOS (with
appropriate $B$-independent functions) and in turn reproduce its shape as 
functions of $B$. As an example, the total TS heat capacity is given by 
$$C_{pTS}(T,B)=\int_0^{\infty}dE~g_{tot}(E,B)C_{p0}(E,T)$$ 
where 
$$C_{p0}(E,T)=k_B\left( \frac{E}{2k_BT} \right)^2\cosh^{-2}\left( 
\frac{E}{2k_BT} \right)$$ 
is the heat capacity contribution from a single TS having energy gap $E$.

We end this Section by mentioning that in model simulation studies of the 
2LSs generated by defects in perfect crystals, Churkin, Barash, and 
Schechter~\cite{CBS2014} found non-uniform behaviour for the $g(E)$ DOS very 
similar to what we advocate in Eq. (\ref{dos}) for the ATS part (see Ref. 
\cite{Jug2004}, Fig. 6 for $B$=0). The (near-) crystallinity is thus the 
common theme between the RERs in glasses and the twin-caged defects in crystals.

A much deeper justification for our ATS model and for the nature of the TSs
in general (of two types only: the 2LSs and the three- or four-fold ATSs) will be 
presented elsewhere, for glasses (papers in preparation).    
  
\section{Heat Capacity}
\subsection{Summary of previous work~\cite{Jug2004}}
In this Section we re-analyze the available data~\cite{Sie2001} for the 
magnetic effect in the heat capacity of two multi-component glasses, 
commercial borosilicate Duran and barium-allumo-silicate (BAS) glass, in order 
to better estimate the concentration of trace Fe-impurities in this way. The 
presence of a magnetic effect in the Pyrex glass was reported long ago by 
Stephens~\cite{Ste1976} and attributed solely to paramagnetic iron impurities 
even though the maximum effect was for $B\simeq0$. A systematic experimental 
study of $C_p(T,B)$ around and below 1 K in some multi-silicate glasses was 
carried out by Siebert~\cite{Sie2001} and those data have been used, upon 
permission, by one of us~\cite{Jug2004} as the very first test of the above 
(Section 2) ETM~\cite{Jug2004}. That earlier analysis best-fitted the $C_p$ 
data by Siebert with the sum of Einstein's $\gamma_{ph}T^3$ phonon term plus 
the 2LS $\gamma_{2LS}T$ non-magnetic contributions, as well as with Langevin's 
paramagnetic and the ATS contributions (see below). The analysis came up with 
concentrations $\bar{n}_J\simeq$48 ppm and, respectively, $\bar{n}_J\simeq$ 20 
ppm instead of the quoted~\cite{Sie2001} 126 ppm (or 180 ppm in a different 
study~\cite{Her2000}) and 102 ppm for Duran and for BAS glass, respectively. 

In order to better understand this large discrepancy we begin with by 
re-analysing Siebert's data for $C_p(T,B)$, after subtraction of the data 
taken at the same temperatures for the same glass, but in the presence of the 
strongest applied magnetic field (8~T). In this way only the magnetic-field 
dependent contributions should remain in the data for 
$\bar{C}_p(T,B)\equiv C_p(T,B)-C_p(T,\infty)$. It should be remarked that 
use of these data for $C_p$ is marred by the question of the many relaxation 
times in a glass, since the data are obtained indirectly (like in all $C_p$ 
measurements in glasses at low $T$), from a heat-pulse experiment where the 
change with time of the sample's temperature is fitted with a single 
relaxation time involving $C_p$. The data thus obtained for $C_p$, however, 
strongly resemble those obtained earlier on by Stephens 
(at $B$=0 and 3.3 T only)~\cite{Ste1976} and there is theoretical 
evidence that the width of the relaxation times' distribution narrows 
considerably in a magnetic field for a multi-component glass~\cite{Jug2015}. 
Still, we use the data of \cite{Sie2001} tentatively.             

\begin{figure}[!hbtp]
\centering
{
   \subfigure[]{\includegraphics[scale=0.60] {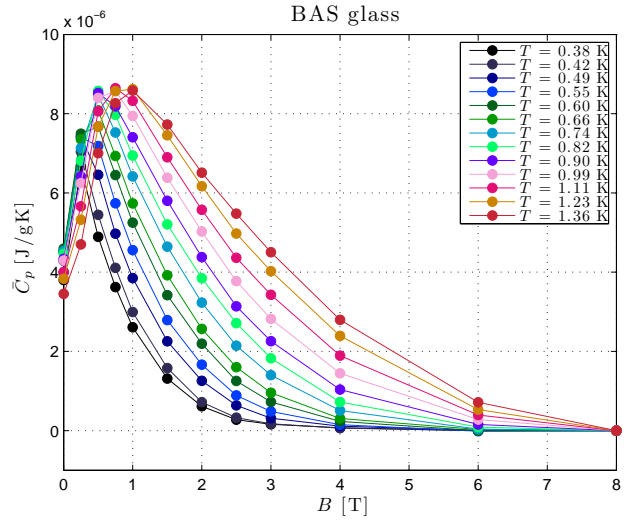} \label{alba_cp_vs_B}}
   \subfigure[]{\includegraphics[scale=0.60] {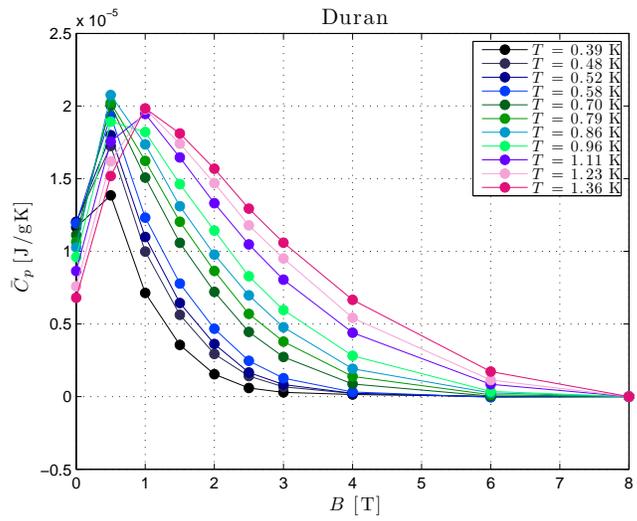} \label{duran_cp_vs_B}} 
 }     
\caption{The heat capacity data $\bar{C}_p=C_p-C_p$(8~T) as a function of the 
magnetic field $B$ for different temperatures: a) for the BAS and b) Duran 
glasses. Data from \cite{Sie2001} (also reproduced, upon permission, in 
\cite{Jug2004}).}
\label{heat_capa}
\end{figure}

Fig.~\ref{heat_capa} presents the data for the heat capacity after subtraction 
of the magnetic field independent data at $B$=8~T for the BAS glass 
(Fig.~\ref{alba_cp_vs_B}) and for Duran (Fig.~\ref{duran_cp_vs_B}). The 
parameters to be determined are the cutoff $D_{min}$ and combinations of 
cutoffs, charge and area $D_{0min}qS$ and $D_{0max}qS$~\cite{Jug2004}, as 
well as:
\begin{center}
\begin{tabular}{ll}
\textbf{$n_{Fe^{2+}}$} & Fe$^{2+}$ impurity concentration\\
\textbf{$n_{Fe^{3+}}$} & Fe$^{3+}$ impurity concentration\\
\textbf{$n_{ATS}$} & ATS concentration (always multiplied by $P^\ast$) \\
\label{tab_imp_cost}
\end{tabular}
\end{center}
The data of Fig.~\ref{alba_cp_vs_B} and Fig.~\ref{duran_cp_vs_B} (we have
restricted the best fit to the three temperatures having the most data points 
around the peak of $C_p(B)$) have been best-fitted by using the following 
magnetic-dependent contributions:
\begin{enumerate}
\item the known Langevin contribution of the paramagnetic Fe impurities 
(Fe$^{2+}$ and Fe$^{3+}$) having concentration $n_J$:
\begin{equation}
C_J(T,B)=n_J \frac{k_B z^2}{4} \bigg(\bigg(\frac{1}{\sinh \frac{z}{2}}\bigg)^2-\bigg(\frac{2J+1}{\sinh \frac{(2J+1)z}{2}}\bigg)^2\bigg)
\label{param_impu_formula_cp}
\end{equation}
where $z=\frac{g\mu_B JB}{k_B T}$ and where $g$ is Land\`e's factor for the
paramagnetic ion in that medium, $\mu_B$ is Bohr's magneton and $J$ the ion's 
total angular momentum (in units $\hbar$=1); $k_B$ is Boltzmann's constant. 
We have assumed the same values the parameters $g$ and $J$ take for Fe$^{2+}$ 
and Fe$^{3+}$ in crystalline SiO$_2$: $J$=2 with $g$=2 and, respectively, 
$J=5/2$ with $g$=2 (we have adopted, in other words, complete quenching of 
the orbital angular momentum~\cite{Abr1961}, consistent with other Authors' 
analyses~\cite{Sie2001,Lud2003}).   
\item the averaged contribution of the ATSs~\cite{Jug2004}, written in 
terms of a sum of individual contributions from each ATS of lowest energy 
gap $E$ 
\begin{equation}
\begin{split}
&C_{ATS}(T,\varphi)=\frac{\pi}{4} \frac{P^{\ast}~n_{ATS}}{k_BT^2}\\ 
&\times\Big\lbrace
\int_{E_{c1}}^{E_{c2}}~dE~\frac{E}{\cosh^2(\frac{E}{2k_BT})}~
\ln \Big[ \frac{(E^2-D_{0min}^2\varphi^2)(E^2-D_{min}^2)}
{D_{min}^2D_{0min}^2\varphi^2} \Big] \\
&+\int_{E_{c2}}^{\infty}~dE~\frac{E}{\cosh^2(\frac{E}{2k_BT})}~
\ln \Big[ \big( \frac{D_{0max}}{D_{0min}} \big)^2 
\frac{E^2-D_{0min}^2\varphi^2}{E^2-D_{0max}^2\varphi^2} \Big]
\Big\rbrace\\
\end{split}
\end{equation}
or, re-written in a dimensionless form as
\begin{equation}
\begin{split}
&C_{ATS}(T,\varphi)=\widetilde{C}_0(T,\varphi)+2\pi P^{\ast} n_{ATS} k_B \Big\lbrace\big[I(x_{c1})-I(x_{c2})\big]\ln(x_{min}x_{0min}\varphi)\\
&+\frac{1}{2}\big[\mathcal{I}(x_{c1},x_{min})-\mathcal{I}(x_{c2},x_{min})+\mathcal{I}(x_{c1},x_{0min}\varphi)-\mathcal{I}(x_{c2},x_{0max}\varphi)\Big\rbrace\\
\end{split}
\label{heat_capa_tot_formula}
\end{equation}
where: 
\begin{itemize}
\item $E_{c1}=\sqrt{D_{min}^2+D_{0min}^2 \varphi^2}$ and $E_{c2}=\sqrt{D_{min}^2+D_{0max}^2 \varphi^2}$;  
\item $x_{c1,2}=\frac{E_{c1,2}}{2k_BT}$, $x_{min}=\frac{D_{min}}{2k_BT}$, etc.; 
\item $I(x) \equiv x \tanh x-\ln \cosh x$;
\item $\mathcal{I}(x,a) \equiv \int_x^\infty \mathrm{d}y \frac{y}{\cosh^2 y}\ln(y^2-a^2)$.
\end{itemize}
and the following expression:
\begin{equation}
\widetilde{C}_0(T,\varphi)=2\pi P^{\ast} n_{ATS} k_B \ln\bigg(\frac{D_{0max}}{D_{0min}}\bigg)\big\lbrace \ln(2)-I(x_{c2}). \big\rbrace\\
\end{equation}
The angular average over the ATS orientations is performed by replacing 
$\varphi\rightarrow\frac{\varphi}{\sqrt{3}}$ (in other words averaging 
$\cos^2\theta$,
$\theta$ being the orientation of ${\bf S}$ with respect to ${\bf B}$).
\end{enumerate}

\subsection{Extracted parameters from best-fitting the heat capacity data}
\subsubsection{BAS glass}
The concentrations of the ATSs and Fe-impurities extracted from the best fit of 
the heat capacity as a function of $B$, for the BAS glass, are reported in 
Table~\ref{tab_imp_extr}; having fixed the concentrations, it was possible to 
extract the other parameters for the BAS glass (Table~\ref{tab_d0_ALBASI_extr}).
The best fit of the chosen data is reported in Fig.~\ref{c_tot_albasi}. 
\begin{table}[!h]
\begin{center}
\begin{tabular}{|c|c|c|}
\hline
BAS glass & Concentration $\mathrm{[g^{-1}]}$ & Concentration [ppm] \\
\hline
\hline
\textbf{$n_{Fe^{2+}}$} & 1.06$\times10^{17}$ & 14.23 \\
\textbf{$n_{Fe^{3+}}$} & 5.00$\times10^{16}$ &  6.69\\
\textbf{$P^{\ast}n_{ATS}$} & 5.19$\times10^{16}$ & - \\
\hline
\end{tabular}
\caption{Extracted parameters (from the heat capacity data) for the 
concentrations of ATSs and Fe-impurities for the BAS glass.}
\label{tab_imp_extr}
\end{center}
\end{table}
\begin{table}[!h]
\begin{center}
\begin{tabular}{|c|c|c|c|}
\hline
Temperature [K] & $D_{min}$ [K] & $D_{0min}\vert\frac{q}{e}\vert S$ [K$\AA^2$] & $D_{0max}\vert\frac{q}{e}\vert S$ [K$\AA^2$]\\
\hline
\hline
0.60 & 0.49 & 4.77$\times 10^{4}$ & 3.09$\times 10^{5}$ \\
0.90 & 0.53 & 5.07$\times 10^{4}$ & 2.90$\times 10^{5}$ \\
1.36 & 0.55 & 5.95$\times 10^{4}$ & 2.61$\times 10^{5}$ \\
\hline
\end{tabular}
\caption{Extracted tunneling parameters (from the $C_p$ data) for the BAS 
glass.}
\label{tab_d0_ALBASI_extr}
\end{center}
\end{table}
\subsubsection{Duran}
\begin{figure}[!Htbp]
\centering
{
  \subfigure[]{\includegraphics[scale=0.60] {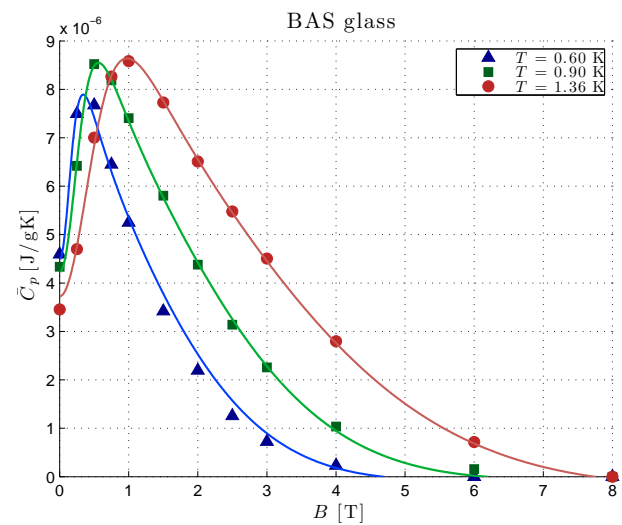} \label{c_tot_albasi}}
  \subfigure[]{\includegraphics[scale=0.60] {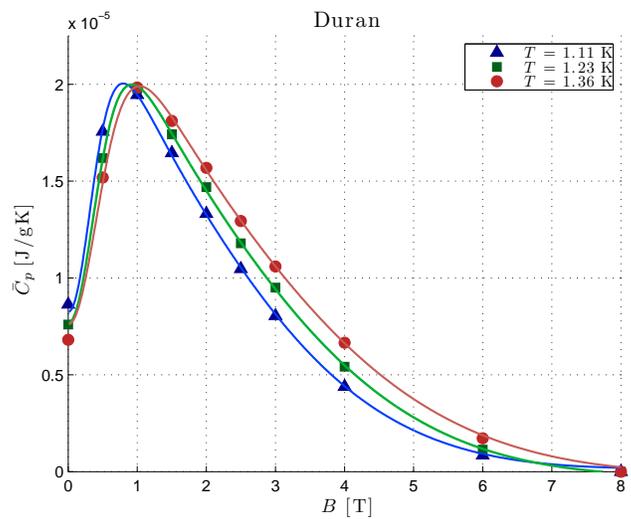} \label{c_tot_duran} }
 } 
\caption{The heat capacity best fit for the a)~BAS and b)~Duran glasses.}
\label{heat_capa_fit}
\end{figure}

The concentrations of the ATSs and Fe-impurities extracted from the best fit of
the heat capacity as a function of $B$, for Duran, are reported in
Table~\ref{tab_imp_extr_dur}; having fixed the concentrations, it was possible
to extract the other parameters for Duran (Table~\ref{tab_d0_duran_extr}). The
fit of the chosen data is reported in Fig.~\ref{c_tot_duran}.
The first comment we make is that these good fits, with a smaller set of 
fitting parameters, repropose concentrations and tunneling parameters very 
much in agreement with those previously obtained by one of us~\cite{Jug2004}. 
The problem with 
the concentrations of the Fe-impurities reported in the literature is that they
do not allow for a good fit of the $\bar{C}_p(B)=C_p(B)-C_p(\infty)$ data in 
the small field region when the Langevin contribution alone is employed 
(Eq.~(\ref{param_impu_formula_cp}) without Eq.~(\ref{heat_capa_tot_formula})).
See for instance Section 5, Fig.~\ref{spec_heat_contri_duran}. 
The Langevin contribution drops to zero below the peak, whilst both Siebert's 
and Stephens' data definitely point to a non-zero value of 
$\bar{C}_p(0)=C_p(0)-C_p(\infty)$ at $B=0$ for any $T>0$. This non-zero 
difference is well accounted for by our ETM and it comes from the ATS 
contribution to the DOS in Eq.~(\ref{dos}).    
\begin{table}[!h]
\begin{center}
\begin{tabular}{|c|c|c|}
\hline
Duran & Concentration $\mathrm{[g^{-1}]}$ & Concentration [ppm] \\
\hline
\hline
\textbf{$n_{Fe^{2+}}$} & 3.21$\times10^{17}$ &  33.01 \\
\textbf{$n_{Fe^{3+}}$} & 2.11$\times10^{17}$ &  21.63 \\
\textbf{$P^{\ast}n_{ATS}$} & 8.88$\times10^{16}$ & - \\
\hline
\end{tabular}
\caption{Extracted parameters (from the heat capacity data) for the
concentration of ATSs and Fe-impurities for Duran.}
\label{tab_imp_extr_dur}
\end{center}
\end{table}
\begin{table}[!h]
\begin{center}
\begin{tabular}{|c|c|c|c|}
\hline
Temperature [K] & $D_{min}$ [K] & $D_{0min}\vert\frac{q}{e}\vert S$ [K$\AA^2$] & $D_{0max}\vert\frac{q}{e}\vert S$ [K$\AA^2$]\\
\hline
\hline
1.11 & 0.34 & 4.99$\times 10^{4}$ & 2.68$\times 10^{5}$ \\
1.23 & 0.32 & 5.30$\times 10^{4}$ & 2.50$\times 10^{5}$ \\
1.36 & 0.32 & 5.54$\times 10^{4}$ & 2.46$\times 10^{5}$ \\
\hline
\end{tabular}
\caption{Extracted tunneling parameters (from the $C_p$~data) for Duran.}
\label{tab_d0_duran_extr}
\end{center}
\end{table}

The results of our $C_p$ analysis definitely indicate that the concentration of 
paramagnetic impurities in the multi-silicate glasses is much lower than 
previously thought and extracted from SQUID-magnetometry measurements of the 
magnetization $M(T,B)$ at moderate to strong field values and as a function of 
$T$. Therefore we now turn our attention to a re-analysis of the 
SQUID-magnetometry data.

\section{Magnetization}
\subsection{Theory}
The second comment we make, having obtained a qualitatively (and also 
quantitatively, barring the multiple relaxation-times issue) good fit to the 
$C_p(B)$ data with the ATS contribution added to the Langevin's, is that the 
ATSs now appear to carry considerably high magnetic moments $\mu_{ATS}$. 
Estimating from the definition ($T=0$)
$\mu_{ATS}= -\frac{\partial}{\partial B}\big(-\frac{1}{2}E\big)$, 
where $E=\sqrt{D^2+D_0^2\varphi^2}$ is the ATS lower energy gap, we get for 
not too small fields $B$ ($\mu_{ATS}$ vanishes linearly with $B$ when 
$B\to 0$, but saturates at high enough $B$):
\begin{equation}
\mu_{ATS}\simeq\frac{\pi}{\Phi_0}SD_0=\frac{\pi}{\phi_0}\left( 
\big\vert\frac{q}{e}\big\vert SD_0 \right).  
\label{atsmoment}
\end{equation}
Thus the very same combination $\frac{q}{e} SD_0$ of parameters appears, 
whilst $\phi_0\equiv h/e$ is the electronic magnetic flux quantum. Using the 
values extracted from the $C_p$ best fit (e.g. Table~\ref{tab_d0_duran_extr}) 
we deduce from Eq.~(\ref{atsmoment}) that (for Duran) $\mu_{ATS}$ ranges from 
about 3.8$\mu_B$ to 27.1$\mu_B$. This fact alone indicates that a large group 
of correlated charged atomic particles is involved in each single ATS and that 
an important ATS contribution to the sample's magnetization is to be expected
(Fe$^{2+}$ and Fe$^{3+}$ have magnetic moment $\mu_J=4\mu_B$ and,
respectively, $5\mu_B$). 
  
The magnetization~$M$ of a sample containing dilute paramagnetic impurities as
well as dilute magnetic-field sensitive ATSs is, like $C_p$, also given by the 
sum of two different contributions:
\begin{enumerate}
\item Langevin's well-known paramagnetic impurities' contribution (Fe$^{2+}$ 
and Fe$^{3+}$, with $n_J$ concentration of one species having spin $J$), given 
by the standard expression~\cite{AM} 
\begin{equation}
M_J=n_J g \mu_B J B_J(z), ~~~\bigg(z=\frac{g\mu_B BJ}{k_BT}\bigg)
\label{magnet_impur}
\end{equation}
where the Brillouin function~$B_J$ is defined by:
\begin{equation}
B_J(z)=\frac{2J+1}{2J}\coth\bigg(\frac{(2J+1)}{2J}z\bigg)-\frac{1}{2J}\coth\bigg(\frac{1}{2J}z\bigg)
\end{equation}
and its low-field susceptibility is the known Curie law:
\begin{equation}
\frac{M}{B}\cong\frac{n_Jg^2\mu_B^2J(J+1)}{3k_BT}
\end{equation}
\item the ATS tunneling currents' contribution, given by the following novel 
expression as the sum of contributions from ATSs of lowest gap $E$: 
\begin{equation}
\begin{split}
M_{ATS}=\pi~P^{\ast} n_{ATS} \frac{1}{B} \bigg\lbrace &\int_{E_{c1}}^{E_{c2}} \mathrm{d}E \tanh\bigg(\frac{E}{2k_BT}\bigg)\ln \bigg(\frac{E^2-D_{0min}^2\varphi^2}{D_{min}^2}\bigg) \\
+&\int_{E_{c2}}^{\infty} \mathrm{d}E  \tanh\bigg(\frac{E}{2k_BT}\bigg)\ln \bigg(\frac{E^2-D_{0min}^2\varphi^2}{E^2-D_{0max}^2\varphi^2}\bigg) \bigg\rbrace
\end{split}
\end{equation}
and which can be also re-expressed (like in the case of $C_{ATS}$) using 
$y=\frac{E}{2k_BT}$ in the following form:
\begin{equation}
\begin{split}
M_{ATS}=2\pi~P^{\ast} n_{ATS} k_B T \frac{1}{B} \bigg\lbrace &\int_{x_{c1}}^{x_{c2}} \mathrm{d}y \tanh y \ln \bigg(\frac{y^2-x_{0min}^2\varphi^2}{x_{min}^2}\bigg) \\
+&\int_{x_{c2}}^{\infty} \mathrm{d}y  \tanh y \ln \bigg(\frac{y^2-x_{0min}^2\varphi^2}{y^2-x_{0max}^2\varphi^2}\bigg) \bigg\rbrace
\end{split}
\label{magnet_adim}
\end{equation}
with, as before:
\begin{itemize}
\item $E_{c1}=\sqrt{D_{min}^2+D_{0min}^2 \varphi^2}$ and $E_{c2}=\sqrt{D_{min}^2+D_{0max}^2 \varphi^2}$;  
\item $x_{c1,2}=\frac{E_{c1,2}}{2k_BT}$, $x_{min}=\frac{D_{min}}{2k_BT}$, etc.; 
\end{itemize}
\end{enumerate}
We present this expression here for the first time, also motivated by the fact 
that we expect a contribution to the measured magnetization $M$ from the ATSs 
that is comparable to, or even greater than, Langevin's paramagnetism of the 
diluted Fe impurities. The above expression follows from a straightforward 
application of standard quantum statistical mechanics, with 
$${\bf M}_{ATS}=n_{ATS}\langle -\frac{\partial {\cal H}_{3LS}}{\partial {\bf B}}
\rangle,$$
$n_{ATS}$ being the ATSs' concentration (a parameter always lumped together
with $P^{\ast}$) and with ${\cal H}_{3LS}$ given by Eq.~(\ref{3lsmagtunneling}).
The angular brackets $\langle \cdots \rangle$ denote quantum, statistical and 
disorder averaging.

The above formula for $M_{ATS}$ is in fact correct for weak magnetic fields. 
For higher fields a correction has to be introduced, owing to the fact that an
improved analytic expression~\cite{Pal2011} for the lowest ATS energy gap must 
be used:  
$$E=\sqrt{D^2+D_0^2\varphi^2(1-\frac{1}{27}\varphi^2)}.$$ 
A full derivation and the study of the $B$- and $T$-dependence of $M_{ATS}$ 
will be presented elsewhere.

\subsection{Extracted parameters for the magnetization data}
\subsubsection{BAS glass}
\begin{figure}[!h]
  \centering
  \includegraphics[scale=0.60] {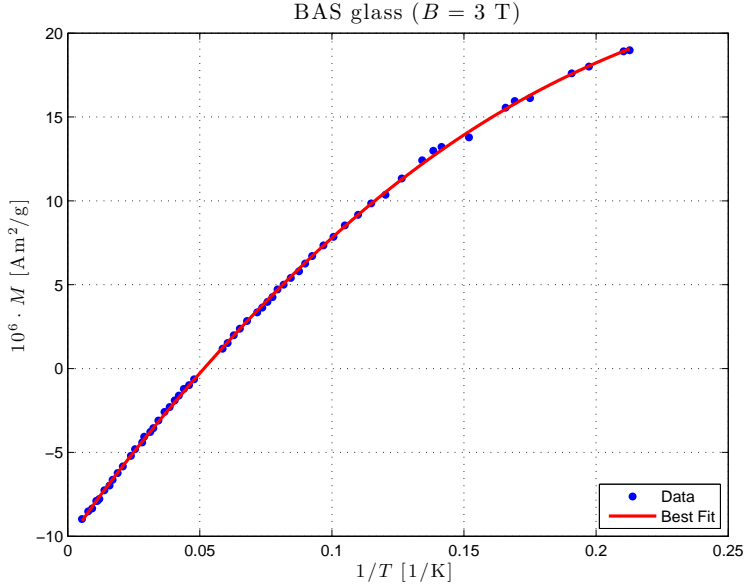} 
\caption{The best fit of the magnetization data~\cite{Sie2001} for the BAS glass, using Eq.~(\ref{magnet_impur}) (for the Fe$^{2+}$ and Fe$^{3+}$ impurities) and Eq.~(\ref{magnet_adim}) (for the ATSs).}
  \label{magnetizat_albasi}
\end{figure}

The magnetization data~\cite{Sie2001} were best-fitted with 
Eq.~(\ref{magnet_impur}) (for the Fe$^{2+}$ and Fe$^{3+}$ contributions) as 
well as with Eq.~(\ref{magnet_adim}) (for the ATSs'), using the parameters
from the $C_p$-fits as input. 
The best fit for the BAS glass is reported in Fig.~\ref{magnetizat_albasi} and 
the extracted parameters in Table~\ref{tabular_all_albasi_magn}.
\begin{table}[!h]
\begin{center}
\begin{tabular}{|c|c|}
\hline
Parameter & BAS glass \\
\hline
\hline
\textbf{$n_{Fe^{2+}}$} $\mathrm{[g^{-1}]}$ & 1.08$\times10^{17}$  \\
\textbf{$n_{Fe^{3+}}$} $\mathrm{[g^{-1}]}$ & 5.01$\times10^{16}$ \\
\textbf{$P^{\ast}n_{ATS}$} $\mathrm{[g^{-1}]}$ & 5.74$\times10^{16}$ \\
$D_{min}$ [K]  & 8.01$\times10^{-2}$\\
$D_{0min}\vert\frac{q}{e}\vert S$ [K$\AA^2$]  & 1.31$\times10^{5}$\\
$D_{0max}\vert\frac{q}{e}\vert S$ [K$\AA^2$] & 2.44$\times10^{5}$\\
$\mathrm{vert. offset}$ [Am$^2$g$^{-1}$] & -1.04$\times10^{-5}$\\
\hline
\end{tabular}
\caption{Extracted parameters (from the magnetization data of \cite{Sie2001}) 
for the concentration of ATS and Fe-impurities of the BAS glass. The vertical 
offset represents the residual Larmor diamagnetic contribution.}
\label{tabular_all_albasi_magn}
\end{center}
\end{table}
\subsubsection{Duran}
\begin{figure}[!h]
  \centering
  \includegraphics[scale=0.60] {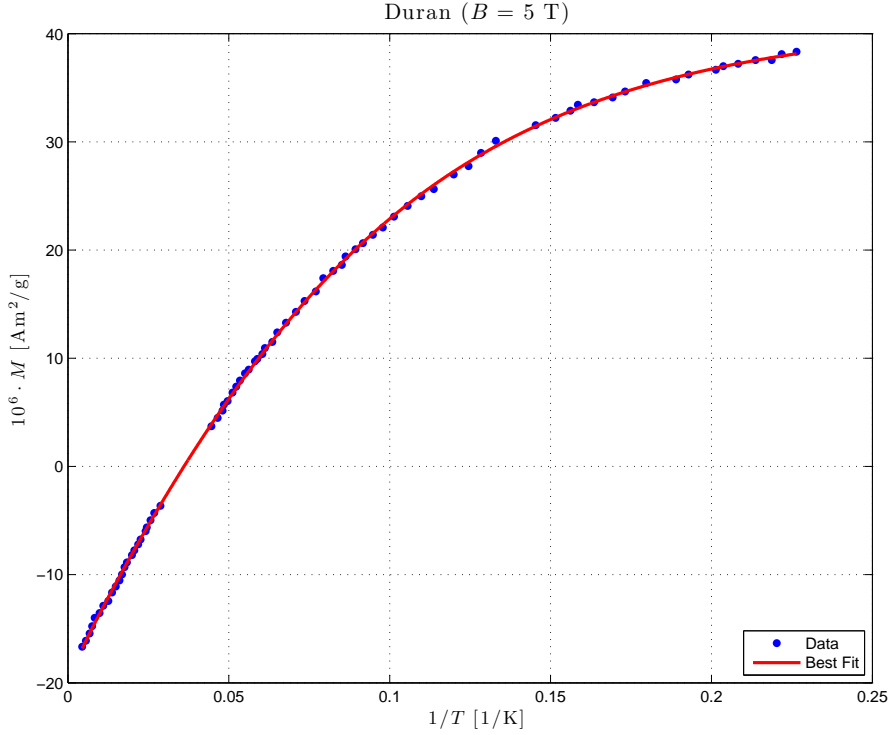} 
\caption{The best fit of the magnetization data~\cite{Sie2001} of Duran, using Eq.~(\ref{magnet_impur}) (for the Fe$^{2+}$ and Fe$^{3+}$ impurities) and Eq.~(\ref{magnet_adim}) (for the ATSs).}
  \label{magnetizat_duran}
\end{figure}

The best fit for Duran is reported in Fig.~\ref{magnetizat_duran} and the 
extracted parameters in Table~\ref{tabular_all_duran_magn}.
\begin{table}[!h]
\begin{center}
\begin{tabular}{|c|c|}
\hline
Parameter & Duran \\
\hline
\hline
\textbf{$n_{Fe^{2+}}$} $\mathrm{[g^{-1}]}$ & 3.07$\times10^{17}$  \\
\textbf{$n_{Fe^{3+}}$} $\mathrm{[g^{-1}]}$ & 2.13$\times10^{17}$ \\
\textbf{$P^{\ast}n_{ATS}$} $\mathrm{[g^{-1}]}$ & 8.68$\times10^{16}$ \\
$D_{min}$ [K]  & 5.35$\times10^{-2}$\\
$D_{0min}\vert\frac{q}{e}\vert S$ [K$\AA^2$]  & 2.00 $\times10^{5}$\\
$D_{0max}\vert\frac{q}{e}\vert S$ [K$\AA^2$] & 2.81$\times10^{5}$\\
$\mathrm{vert. offset}$ [Am$^2$g$^{-1}$] & -1.97$\times10^{-5}$\\
\hline
\end{tabular}
\caption{Extracted parameters (from the magnetization data of \cite{Sie2001}) 
for the concentration of ATS and Fe impurities of Duran.}
\label{tabular_all_duran_magn}
\end{center}
\end{table}
\subsection{Concentration conversion}
We now explain how we convert the Fe-concentrations thus obtained to 
atomic ppm concentrations (ppma). $n_J$, the mass density of a Fe species with
spin $J$ in the sample, is given by:
\begin{equation}
n_J=\frac{N_J}{M}=\frac{N_J}{N_{at}}\frac{N_A}{\sum_i \xi_i A_i}
\label{n_JFE}
\end{equation}
where
\begin{equation} 
M=\sum_i \xi_i\frac{N_{at}}{N_A}A_i 
\end{equation}
is the sample's mass, and where:
\begin{center}
\begin{tabular}{ll}
$N_J$ &  number of Fe-ions in the sample with spin $J$ \\
$\xi_i$ & molar fraction of the i-$th$ species \\
$A_i$ & molar mass of the i-$th$ species \\
$N_{at}$ & total number of atoms in the sample \\
$N_A$ & Avogadro's number (6.022$\times$10$^{23}$~mol$^{-1}$) \\
\end{tabular}
\end{center}
for the Fe$^{2+}$ ($J$=2) and Fe$^{3+}$ ($J$=5/2) impurities.  
Table~\ref{bas_duran_chem_compo} shows parameters related to the chemical and
molar composition~\cite{Sie2001} of the two multi-component silicate
glasses. Therefore, using the parameters reported in 
Table~\ref{bas_duran_chem_compo}, one has:
$$ 
\sum_i \xi_i A_i = 
\left\{
\begin{array}{ll}
80.530 \mathrm{\frac{g}{mol}} &\mathrm{for~BAS~glass} \\
61.873 \mathrm{\frac{g}{mol}} &\mathrm{for~Duran} \\
\end{array}
\right.
$$
\begin{table}[!Htp]
\begin{center}
\begin{tabular}{|c|c|c|c|}
\hline
Oxide Element & Molar Mass ($A_i$) & $\xi_{i,\%}$ BAS glass & $\xi_{i,\%}$ Duran \\
\hline
SiO$_2$ & 60.084 & 72.7 & 83.4 \\
\hline
B$_2$O$_3$ & 69.620 & 0.72 & 11.6 \\
\hline
Al$_2$O$_3$ & 101.961 & 8.8 & 1.14\\
\hline
Na$_2$O & 61.979 & 0.28 & 3.4 \\
\hline
K$_2$O & 94.196 & 0.064 & 0.41\\
\hline
BaO & 153.326 & 17.0 & 0.005\\
\hline
Li$_2$O & 29.881 & 0.014 & 0.004 \\
\hline
PbO & 223.199 & 0.48 & $<$0.01 \\
\hline
\end{tabular}
\end{center}
\caption[]{Molar mass~$A_i$ and percentage fraction~$\xi_{i,\%}$ of the various oxides making up the BAS (third column) and Duran (fourth column) glasses, as reported in~\cite{Sie2001,Lud2003}.}
\label{bas_duran_chem_compo}
\end{table}
We interpret $N_J/N_{at}\equiv\bar{n}_J$ as the atomic concentration of the 
spin-$J$ Fe species (to be multiplied by 10$^6$ to obtain the ppm) and 
thus we have the conversion formula:
\begin{equation}
n_J=\bar{n}_J\frac{N_A}{\sum_i \xi_i A_i}
\label{conversion}
\end{equation}
\section{Conclusions: Extracted Concentrations for Iron Impurities and ATSs}
\subsection{BAS glass}
The nominal concentration of Fe$^{3+}$ for the BAS glass is (using Eq.~(\ref{n_JFE}) or (\ref{conversion})):
\begin{equation}
\begin{split}
\bar{n}_{Fe^{3+}_{nom}}&=102~\mathrm{ppm} \\
n_{Fe^{3+}_{nom}}&=\frac{10^{-6}\cdot 102 \cdot 6.022\times10^{23}~\mathrm{mol}^{-1}}{80.530 \mathrm{\frac{g}{mol}}}=7.63\cdot 10^{17} \mathrm{g^{-1}}\\
\end{split}
\end{equation}
which is inadequate (as in Duran's case below) to explain the behaviour of the 
heat capacity as a function of $B$, here presented (Fig.~\ref{heat_capa_fit}) 
and as a function of $T$ (studied in~\cite{Jug2004}).
Table~\ref{tab_imp_extr_alba_tot} summarizes the concentrations found from 
our best fits of heat capacity and magnetization data, for the BAS glass.
The parameters found in~\cite{Jug2004} were 
$P^{\ast}n_{ATS}$=6.39$\times10^{16}$~g$^{-1}$ and $\bar{n}_{Fe}$=20.44~ppm 
where this latter was for the Fe$^{2+}$ concentration only.
The present study confirms that most of the Fe-impurities in these two glasses
are of the Fe$^{2+}$ type~\cite{Jug2004}.
\begin{table}[!h]
\begin{center}
\begin{tabular}{ |l|l| }
  \hline  
  \multicolumn{2}{|c|}{{BAS glass}} \\
  \hline
  \multicolumn{2}{|c|}{Heat Capacity fit} \\
  \hline
  \textbf{$n_{Fe^{2+}}$} & 1.06$\times10^{17}~\mathrm{g^{-1}}$ = 14.23~ppm \\
  \textbf{$n_{Fe^{3+}}$} & 5.00$\times10^{16}~\mathrm{g^{-1}}$ =  6.69~ppm \\
  \textbf{$P^{\ast}n_{ATS}$} & 5.19$\times10^{16}~\mathrm{g^{-1}}$  \\ 
  \hline
  \hline
  \multicolumn{2}{|c|}{Magnetization fit} \\
  \hline
  \textbf{$n_{Fe^{2+}}$} & 1.08$\times10^{17}~\mathrm{g^{-1}}$ = 14.38~ppm  \\
  \textbf{$n_{Fe^{3+}}$} & 5.01$\times10^{16}~\mathrm{g^{-1}}$ = 6.70~ppm\\
  \textbf{$P^{\ast}n_{ATS}$} & 5.74$\times10^{16}~\mathrm{g^{-1}}$ \\
   \hline
\end{tabular}
\caption{Comparison between the concentrations extracted from the two different 
best-fitted experimental data sets for the BAS glass.}
 \label{tab_imp_extr_alba_tot}
\end{center}
\end{table}
\subsection{Duran}
\begin{figure}[!Htp]
  \centering
  \includegraphics[scale=0.60] {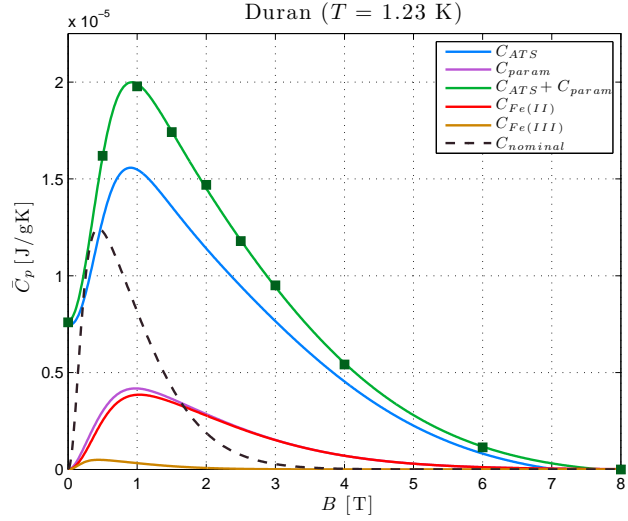} 
\caption{The curves represent the different terms that contribute to the heat 
capacity of Duran in our best fit of the data from \cite{Sie2001}. The dashed 
curve is for Langevin's contribution only, but with the nominal concentration 
of $\bar{n}_{Fe^{3+}}$=126 ppm (no ATS). }
  \label{spec_heat_contri_duran}
\end{figure}

The nominal concentration of Fe$^{3+}$ for Duran is (using Eq.~(\ref{n_JFE}) or (\ref{conversion})):
\begin{equation}
\begin{split}
\bar{n}_{Fe^{3+}_{nom}}&=126~\mathrm{ppm} \\
n_{Fe^{3+}_{nom}}&=\frac{10^{-6}\cdot 126 \cdot 6.022\times10^{23}~\mathrm{mol}^{-1}}{61.873 \mathrm{\frac{g}{mol}}}=1.23\cdot 10^{18} \mathrm{g^{-1}}\\
\end{split}
\end{equation}
which again is inadequate to explain the behaviour of the heat capacity as a 
function of $B$ (see Fig.~\ref{spec_heat_contri_duran}). 
Table~\ref{tab_imp_extr_duran_tot} summarizes the concentrations found from 
our best fits of heat capacity and magnetization data, for Duran. The 
parameters found in~\cite{Jug2004} were 
$P^{\ast}n_{ATS}$=6.92$\times10^{16}$~g$^{-1}$ and $\bar{n}_{Fe}$=47.62~ppm 
where this latter was the Fe$^{2+}$ concentration only.

\begin{figure}[!Htp]
  \centering
  \includegraphics[scale=0.60] {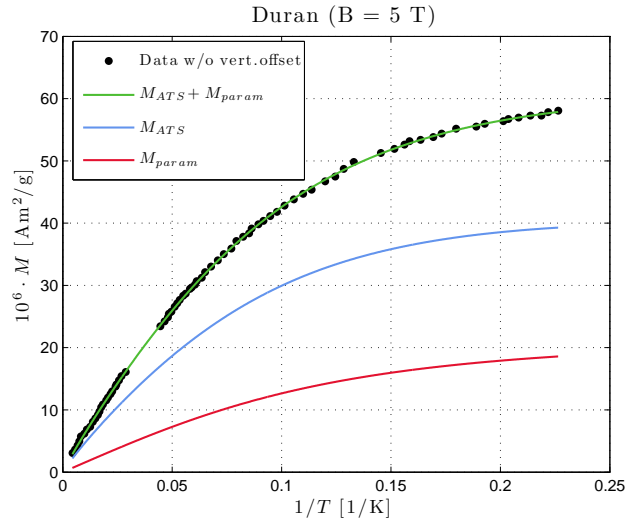} 
\caption{The curves represent the different contributions to the magnetization 
of Duran in our best fit of the data from \cite{Sie2001}.}
  \label{magnetiz_contri_duran}
\end{figure}

Fig.~\ref{spec_heat_contri_duran} presents the behaviour of the different 
contributions to the heat capacity as a function of $B$ for Duran; 
$C_{Fe^{2+}}$ and $C_{Fe^{3+}}$ are given by Eq.~(\ref{param_impu_formula_cp}), 
respectively with the Fe$^{2+}$ and Fe$^{3+}$ parameters, $C_{param}$ is the 
sum of these latter two, C$_{ATS}$ is given by 
Eq.~(\ref{heat_capa_tot_formula}) and the green line represents the result of 
the best fit. The dashed line corresponds to the $\bar{C}_p(B)$ one would get
from the nominal concentration $\bar{n}_{Fe}$ of 126 ppm~\cite{Sie2001} as 
extracted from the SQUID magnetization measurements fitted with the Langevin
contribution only (no ATS contribution). Likewise 
Fig.~\ref{magnetiz_contri_duran} presents the behaviour of the different 
contributions (Eq.(\ref{magnet_impur}) and Eq.(\ref{magnet_adim})) to the 
magnetization as a function of $B$, also for Duran. It can be seen that the ATS 
contribution is in both cases dominant, also (in the case of the magnetization) 
at the higher temperatures.
\begin{table}[!h]
\begin{center}
\begin{tabular}{ |l|l| }
  \hline  
  \multicolumn{2}{|c|}{{Duran}} \\
  \hline
  \multicolumn{2}{|c|}{Heat Capacity fit} \\
  \hline
  \textbf{$n_{Fe^{2+}}$} & 3.21$\times10^{17}~\mathrm{g^{-1}}$ = 33.01~ppm \\
  \textbf{$n_{Fe^{3+}}$} & 2.11$\times10^{17}~\mathrm{g^{-1}}$ =  21.63~ppm \\
  \textbf{$P^{\ast}n_{ATS}$} & 8.88$\times10^{16}~\mathrm{g^{-1}}$  \\ 
  \hline
  \hline
  \multicolumn{2}{|c|}{Magnetization fit} \\
  \hline
  \textbf{$n_{Fe^{2+}}$} & 3.07$\times10^{17}~\mathrm{g^{-1}}$ = 31.58~ppm  \\
  \textbf{$n_{Fe^{3+}}$} & 2.13$\times10^{17}~\mathrm{g^{-1}}$ = 21.86~ppm\\
  \textbf{$P^{\ast}n_{ATS}$} & 8.68$\times10^{16}~\mathrm{g^{-1}}$ \\
   \hline
\end{tabular}
\caption{Comparison between the concentrations extracted from the two different 
best-fitted experimental data sets for Duran.}
 \label{tab_imp_extr_duran_tot}
\end{center}
\end{table}
\subsection{Conclusions and outlook}
In conclusion, by allowing a contribution to the magnetization from the ATS 
tunneling currents as evaluated from our ETM we have been able to show that
the concentrations of Fe impurities (of type Fe$^{2+}$ as well as Fe$^{3+}$) 
are in much better agreement when extracted from SQUID magnetization data or 
from heat capacity data with the samples in a magnetic field. We have aimed 
at a semi-quantitative agreement, but inclusion of a possible contribution to
$C_p$ from Kramers-doublets' splitting for the $J=5/2$ Fe$^{3+}$ 
species~\cite{Sie2001} (which we confirm, however, to be the minority 
one~\cite{Jug2004}) could possibly further improve the quantitative agreement.  

We reach the surprising conclusion that the ATSs and the ETM employed keep 
giving a good description of the glass magnetization till the highest 
temperatures in the SQUID-magnetometry measurements, about 300 K (the 
glass-transition temperature $T_g$ being, however, 1123 K for BAS glass and 
803 K for Duran, respectively~\cite{Sie2001}). This new finding is 
interpreted with the consideration that no phonon-TS interactions are 
involved in the measurement of the magnetization, contrary to the case of AC 
dielectric-constants and polarization-echo measurements, or to the case of 
the acoustic measurements, or that of the heat capacity, where phonons 
do contribute and in a complicated way well above 1 K~\cite{Esq1998,Phi1987}. 
The SQUID magnetization measurement in a magnetic field is thus an ideal 
arena where to study the TSs in general and to test our ETM, proposed for the 
explanation of the magnetic effects in the multi-component glasses as 
manifestations of the inhomogeneous structure of the glassy state. Therefore a
systematic experimental study of SQUID magnetization as a function of $T$ and
$B$ from multi-component glasses with very low paramagnetic impurity 
concentration (e.g. BK7) would be most welcome to test our model. Moreover, 
also the tunneling parameters and the concentration of the ATSs here obtained 
turn out to be similar when extracted from the $C_p$- and from the 
$M$-data, with the tunneling parameters $D_{min}$, 
$D_{0min}\vert\frac{q}{e}\vert S$ and $D_{0max}\vert\frac{q}{e}\vert S$ 
extracted from the $M$-data being as anomalously large as from best-fits of all 
data in the other experiments~\cite{Jug2004,Jug2009,Jug2014,Jug2010}. These 
large values have been interpreted by one of us~\cite{Jug2013} as deriving 
from the correlated tunneling of a large, but not yet mesoscopic or 
macroscopic, number $N\sim$ 200 to 600 of atomic-scale TSs, 
the ATS being only a fictitious tunneling particle involving in fact the 
correlated rearrangement of a large group of (charged) atoms. There appears 
to be a weak temperature dependence $N(T)$ of this number of correlated 
atomic tunnelers, since the parameters quoted above change slightly (or even 
significantly, yet remaining large) from experiment to experiment carried out 
at different temperature ranges. About this revealing $T$-dependence will be 
expanded in future publications.

We remark that the concentration $n_J$ of paramagnetic impurities turns out
to be significantly lower, up to 80\% less, than the concentrations reported
in the literature for the analysed glasses and as extracted from 
SQUID-measurements of the magnetization~\cite{Her2000,Sie2001,Woh2001}. Our 
point of view is that without the inclusion of the contribution from the ATSs 
the extracted SQUID-measurement concentration of paramagnetic impurities will 
be considerably overestimated. This happened already in the case of Stephens' 
data~\cite{Ste1976}, which have led to the Langevin-only 
estimate~\cite{Jug2004} $\bar{n}_{Fe}\simeq$~50 ppm in Pyrex from the $C_p$ 
measurement data in a magnetic field, when in fact the mass-spectrometry 
analysis had given~\cite{Ste1976} $\bar{n}_{Fe}$=12 ppm. The fact that we 
have established, that Langevin-only fitted SQUID-magnetization measurements 
considerably overestimate the concentration of paramagnetic impurities in a 
glassy matrix, would appear to cast serious doubts about the trace 
Fe-impurities and associated paramagnetic-TSs as possible sources of the 
magnetic effects in the cold glasses~\cite{Bor2007}. Indeed, Fe$^{3+}$ would 
enter substitutionally to Si$^{4+}$ only in a crystal (e.g. quartz), whilst 
in a multi-silicate glass the overwhelming majority of Fe-impurities would 
enter as network-modifiers of the SiO$_4$ glassy matrix~\cite{Hen1989}. This 
means that the amplitude of the [FeO$_4$]$^-$ paramagnetic-TS contribution 
should be reduced considerably more than the 80\% we claim from the 
overestimate of the Fe-concentration from Langevin-fitted SQUID-measurements. 
We note in passing that the paramagnetic-TS explanation of the magnetic effect 
in $C_p$ requires concentrations $n_{Fe}$ already some 40\% greater than the 
nominal, Langevin-only SQUID-extracted values~\cite{Bor2007}. The 
paramagnetic-TS approach may nevertheless retain some validity in the case of 
the heavily Fe- and Cr-doped multi-silicate glasses~\cite{Bor2011}. 

\section*{Acknowledgements}
One of us (SB) acknowledges support from the Italian Ministry of Education,
University and Research (MIUR) through a Ph.D. Grant of the Progetto Giovani
(ambito indagine n.7: materiali avanzati (in particolare ceramici) per
applicazioni strutturali), as well as from the Bando VINCI-2014 of the
Universit\`a Italo-Francese. The other Author (GJ) is grateful to the 
Laboratoire des Verres et Collo\"ides in Montpellier for hospitality and for 
many stimulating discussions, as well as to the Referees for useful comments
on the manuscript. Enlightening conversations with Carlo Dossi and Paolo Sala
about glass contaminants are also kindly acknowledged.

\appendix
\renewcommand{\thefigure}{A\arabic{figure}}
\setcounter{figure}{0}
\renewcommand{\thetable}{A\arabic{table}}
\setcounter{table}{0}

\section*{Appendix}
\begin{figure}[!Hbp]
  \centering
  \includegraphics[scale=0.60] {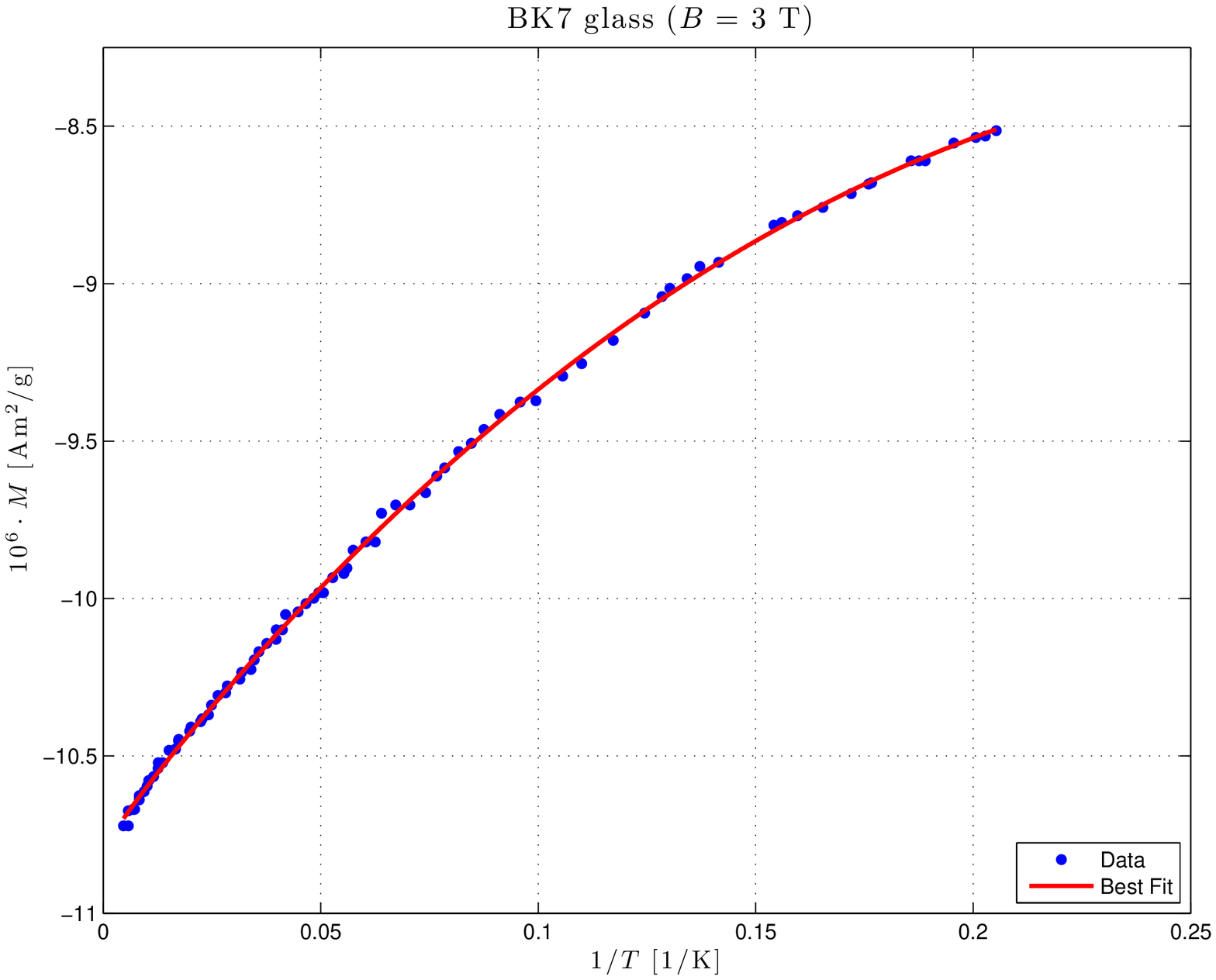} 
\caption{The best fit of the magnetization data~\cite{Sie2001} for BK7, using 
Eq.~(\ref{magnet_impur}) (for the Fe$^{2+}$ and Fe$^{3+}$ impurities) and 
Eq.~(\ref{magnet_adim}) (for the ATSs). Data from \cite{Sie2001}.}
  \label{magnetizat_BK7}
\end{figure}

Here we present our preliminary study of the SQUID magnetization data (also 
available from \cite{Sie2001}) for the borosilicate glass BK7, for which 
however no substantial magnetic effect in the heat capacity $C_p$ has been 
reported~\cite{Sie2001}. This glass has a nominal Fe-impurity concentration 
of $\bar{n}_{Fe^{3+}}$=6 ppm~\cite{Sie2001,Lud2003,Woh2001}, yet our best fit 
in Fig.~\ref{magnetizat_BK7} with both Langevin (Eq.~(\ref{magnet_impur})) and 
ATS (Eq.~(\ref{magnet_adim})) contributions produces the concentrations and 
parameters given in Table~\ref{tabular_all_BK7_magn}. The best fit was carried 
out with knowledge of ATS parameters from our own theory~\cite{Jug2014} for 
the magnetic effect in the polarization-echo experiments at mK 
temperatures~\cite{Lud2003}. We conclude that our main contention is once 
more confirmed, in that the concentration of Fe in BK7 we extract in this way
is only about 1.1 ppm and the bulk of the SQUID magnetization is due to the 
ATSs. Table~\ref{tabular_all_BK7_magn} reports our very first estimate of 
$n_{ATS}P^{\ast}$ for BK7. Assuming $P^{\ast}$ to be of order 1 and about
the same for all glasses, we conclude that the concentration $n_{ATS}$ of the 
ATSs nesting in the RERs is very similar for all of the multi-silicate glasses 
by us studied for their remarkable magnetic effects. From the present 
SQUID-magnetization best fits we have obtained 5.74$\times10^{16}$ g$^{-1}$ 
(BAS glass), 8.68$\times10^{16}$ g$^{-1}$ (Duran) and 1.40$\times10^{16}$
g$^{-1}$ (BK7). The almost negligible magnetic effect in $C_p$ for BK7 is 
due, in our approach, to the low values of the cutoffs $D_{0min}$ and 
$D_{0max}$ for this system (these parameters appearing in the prefactor and
in the integrals' bounds determining the ATS contribution to 
$C_p$~\cite{Jug2013}).  
\begin{table}[!h]
\begin{center}
\begin{tabular}{|c|c|}
\hline
Parameter & BK7 \\
\hline
\hline
\textbf{$n_{Fe^{2+}}$} $\mathrm{[g^{-1}]}$ & 6.69$\times10^{15}$ = 0,71 ppm  \\
\textbf{$n_{Fe^{3+}}$} $\mathrm{[g^{-1}]}$ & 3.43$\times10^{15}$ = 0.36 ppm\\
\textbf{$P^{\ast}n_{ATS}$} $\mathrm{[g^{-1}]}$ & 1.40$\times10^{16}$ \\
$D_{min}$ [K]  & 5.99$\times10^{-2}$\\
$D_{0min}\vert\frac{q}{e}\vert S$ [K$\AA^2$]  & 8.87$\times10^{4}$\\
$D_{0max}\vert\frac{q}{e}\vert S$ [K$\AA^2$] & 1.20$\times10^{5}$\\
$\mathrm{vert. offset}$ [Am$^2$g$^{-1}$] & -1.08$\times10^{-5}$\\
\hline
\end{tabular}
\caption{Extracted parameters (from the magnetization data of \cite{Sie2001}) 
for the concentration of ATSs and Fe impurities of the BK7 ($\sum_i\xi_iA_i$=
63.530 g~mol$^{-1}$~\cite{Sie2001}). The vertical offset represents the 
residual Larmor diamagnetic contribution.}
\label{tabular_all_BK7_magn}
\end{center}
\end{table}

\end{document}